\documentclass[%
 reprint,
 amsmath,amssymb,
 aps,
]{revtex4-2}
\usepackage{hyperref}
\usepackage{comment}
\usepackage{graphicx}% Include figure files
\usepackage{dcolumn}% Align table columns on decimal point
\usepackage{bm}% bold math
\usepackage{xcolor}

\begin{document}

\preprint{APS/123-QED}
\title{Astrophysical origins of TeV features in the cosmic-ray lepton spectrum}
%\title{On the spectrum of astrophysical electrons}% Force line breaks with \\

\author{Zhen Xie}
\affiliation{School of Astronomy and Space Science, University of Science and Technology of China, Hefei, 230026, China}
% --- Ruizhi Yang ---
\author{Ruizhi Yang}
\email{yangrz@ustc.edu.cn}
\affiliation{School of Astronomy and Space Science, University of Science and Technology of China, Hefei, 230026, China}

\date{\today}% It is always \today, today,

\begin{abstract}
Precise measurements of high-energy cosmic-ray electrons and positrons have revealed spectral structures that are difficult to capture with a single smooth power-law background. The rising positron fraction measured by PAMELA and AMS-02, together with the all-electron excess reported by ATIC and the high-precision all-electron spectrum measured by DAMPE, has motivated interpretations ranging from nearby astrophysical accelerators to dark-matter annihilation or decay. In this work we revisit the conventional diffuse electron background and the possible contribution from nearby pulsars in a common propagation framework. The diffuse component is modeled with GALPROP configurations calibrated by cosmic-ray nuclei and diffuse gamma-ray observations. We then use the Green-function solution for nearby discrete sources with radiative losses to study pulsar contributions with both burst-like and continuous injection histories, including the effect of stochastic inverse-Compton cooling on the propagated spectra.
We also use the highest-energy DAMPE data points as an illustrative case to compare possible local-source contributions from pulsars and supernova-remnant-like burst sources.
The spectral shape of such features provides a useful diagnostic for distinguishing physically plausible nearby-source features from more exotic interpretations.
\end{abstract}

\maketitle

%\tableofcontents

\section{Introduction}
High-energy cosmic-ray electrons and positrons provide a sensitive probe of Galactic cosmic-ray propagation, local source activity, and possible contributions from new physics. In the standard picture, charged cosmic rays diffuse through turbulent Galactic magnetic fields and interact with the interstellar medium, radiation fields, and magnetic fields before reaching the Solar system~\cite{Strong2007CRPropagation,Blasi2013OriginCR}. Compared with hadronic cosmic rays, electrons and positrons suffer rapid radiative energy losses through synchrotron emission and inverse-Compton (IC) scattering. Above tens to hundreds of GeV, these losses substantially reduce the propagation horizon, so the observed all-electron spectrum can retain information about nearby and relatively young sources rather than only the fully averaged Galactic source population~\cite{Serpico2012PositronReview,Delahaye2010Positrons}.

The local-source nature of high-energy electrons has been recognized for a long time. Early work pointed out that radiative losses make the continuous-source approximation unreliable at sufficiently high energy and that nearby pulsars could contribute appreciably to the observed electron spectrum~\cite{Shen1970Pulsars}. Direct balloon and emulsion-chamber measurements extending to hundreds of GeV and approaching the TeV scale further motivated this picture~\cite{Nishimura1980Electrons}. Later analytic and numerical studies developed the discrete-source framework more quantitatively, separating the smooth background from the contribution of one or a few nearby accelerators and showing that local sources can generate spectral structures without invoking exotic physics~\cite{Atoyan1995PRD,Aharonian1995AALetter,Kobayashi2004SNR}. Thus the modern question is not simply whether structures can appear in the electron spectrum, but whether their detailed shape and normalization favor a conventional local source or a particle-physics origin.

In recent decades, direct measurements have provided increasingly accurate information on the electron spectrum. PAMELA reported a rising positron fraction above several GeV, in clear tension with the expectation from a purely secondary positron component in conventional propagation models~\cite{Adriani2009PAMELA}. AMS-02 subsequently confirmed this excess with much higher precision and extended the measurement to hundreds of GeV~\cite{Aguilar2013AMSPositron,Aguilar2019AMSPositron}. At the same time, ATIC reported a broad excess in the cosmic-ray electron-plus-positron spectrum at several hundred GeV~\cite{Chang2008ATIC}. Although the detailed ATIC feature was not confirmed as a narrow bump by later measurements, it strongly motivated high-statistics measurements of the all-electron spectrum. Fermi-LAT measured a smooth but relatively hard spectrum up to the TeV scale~\cite{Abdo2009FermiElectron}, H.E.S.S. extended the measurement into the multi-TeV range and found evidence for a steepening~\cite{Aharonian2008HESS,Aharonian2009HESS}, CALET provided an independent calorimetric determination~\cite{Adriani2017CALET}, and DAMPE reported a spectral break near $0.9~{\rm TeV}$~\cite{Ambrosi2017DAMPE}. Together these observations established that the high-energy electron and positron data contain more structure than expected from a single featureless power law.

Generally, these measurements have been interpreted in two broad ways. Dark-matter annihilation or decay into leptonic final states can produce hard electron and positron spectra, while the finite particle mass naturally introduces a cutoff or edge-like feature near the endpoint energy~\cite{Bergstrom2008PAMELA,ArkaniHamed2009DM,Cirelli2009PPPC}. Such interpretations are attractive because they connect charged-particle spectral structures to new particle physics, but they are constrained by gamma-ray observations of the Galactic halo, dwarf spheroidal galaxies, and diffuse emission, by cosmic-microwave-background (CMB) limits on energy injection, and by antiproton measurements that restrict many hadronic channels~\cite{Ackermann2015DwarfDM,Slatyer2016CMB,Aguilar2016AMSAntiproton}. The inferred dark-matter parameters also depend on assumptions about the local density, possible substructure, radiative losses, solar modulation, and the modeling of the conventional electron and positron background.

Astrophysical scenarios provide an alternative explanation. Pulsars and pulsar wind nebulae naturally produce electron-positron pairs in the magnetosphere and wind region and can release hard lepton spectra into the interstellar medium after escape from the nebula~\cite{Gaensler2006PWNReview,Kargaltsev2008PWNe}. Nearby mature pulsars such as Geminga and Monogem have therefore long been studied as possible sources of the positron excess and high-energy all-electron features~\cite{Atoyan1995PRD, Hooper2009Pulsars,Yuksel2009Geminga,Profumo2012Pulsars,Serpico2012PositronReview}. Supernova remnants and other local discrete accelerators can also imprint structures on the electron spectrum, because TeV electrons cool rapidly and probe only a limited volume and time interval of the Galactic source population~\cite{Kobayashi2004SNR,Delahaye2010Positrons}. In this picture, spectral structure can arise from finite source distance, finite age, injection history, and cooling-limited propagation, without requiring exotic particle physics.

These possibilities create a long-standing degeneracy. Dark matter can produce hard leptonic spectra and sharp cutoffs, while nearby astrophysical sources can also generate high-energy features through source cutoffs, discreteness, and ordinary diffusion-loss propagation. Existing data do not automatically distinguish these scenarios. The discriminating power will improve as DAMPE accumulates exposure and as future high-energy-resolution missions such as HERD measure the detailed shape of the all-electron spectrum with larger statistics and high energy resolution at the TeV scale~\cite{Zhang2014HERD}.

In this work we revisit the astrophysical contribution to high-energy cosmic-ray electron spectra, with particular attention to whether nearby sources can produce particle-physics-like structures. We begin in Sec.~\ref{sec:diffuse} by constructing the smooth diffuse background from GALPROP models calibrated by local cosmic-ray measurements and diffuse gamma-ray observations, which provides a physically motivated estimate of the background uncertainty and spectral variation. In Sec.~\ref{sec:pulsar}, we then study nearby discrete-source contributions, focusing mainly on pulsars with burst-like or continuous injection histories and including the effect of stochastic inverse-Compton cooling on the propagated spectra. This analysis separates features inherited from the source injection spectrum from those generated by propagation and cooling, and allows us to examine the corresponding energetic requirements. In Sec.~\ref{sec:tev_component}, we apply this framework to the highest-energy DAMPE data points as an illustrative estimate of the source parameters required for a possible additional TeV-scale component. Finally, Sec.~\ref{sec:discussion} summarizes the implications for distinguishing nearby astrophysical sources from particle-physics interpretations with future high-resolution measurements.

\section{Diffuse cosmic-ray electron background}
\label{sec:diffuse}
A reliable interpretation of high-energy spectral structure first requires a physically defined diffuse background. In many phenomenological studies, the conventional electron component is represented by a power law or a broken power law. Such parameterizations are convenient for fitting the local spectrum, but they do not fully exploit the constraints from cosmic-ray propagation and multi-wavelength electromagnetic  radiation observations, and they can blur the distinction between smooth background curvature and a genuinely additional component. We therefore adopt GALPROP v54 calculations as the baseline description of the diffuse cosmic-ray electron background~\cite{Strong2007CRPropagation,Vladimirov2011GALPROP}. Within this framework, the electron spectrum is linked consistently to the diffusion coefficient, source distribution, radiative losses, interstellar radiation field, and Galactic magnetic field, rather than being treated as an independent empirical curve.

\begin{figure}
    \centering
    \includegraphics[width=1\linewidth]{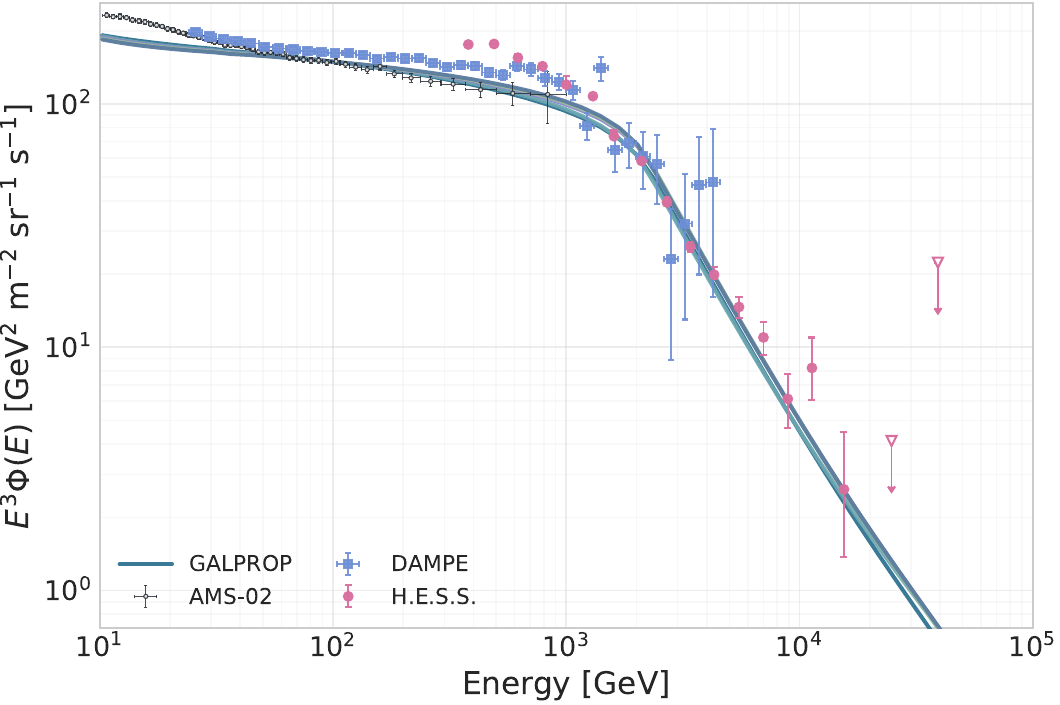}
    \caption{Comparison of representative GALPROP all-electron diffuse-background spectra with AMS-02\cite{Aguilar2014AMSElectron}, DAMPE\cite{Ambrosi2017DAMPE} and H.E.S.S.\cite{HESS2024CRE} measurements. The colored curves show a set of representative GALPROP models selected from the retained propagation-model ensemble to span the range of high-energy spectral shapes. All model curves are normalized to the AMS-02 all-electron flux at $100\,{\rm GeV}$ in order to emphasize differences in spectral shape rather than absolute normalization. The AMS-02 and DAMPE points show the measured all-electron fluxes, while the H.E.S.S. points are the CR-electron-candidate fluxes from Table I of the Supplemental Material of Ref.~\cite{HESS2024CRE}; the two highest-energy H.E.S.S. entries are shown as $99\%$ C.L. upper limits.
}
    \label{fig:eall}
\end{figure}

For the diffuse background, we use the GALPROP parameter set developed in previous Galactic diffuse gamma-ray studies~\cite{Ackermann2012Diffuse}. In particular, the parameter set spans variations in the cosmic-ray halo height, source distribution, and gas density profile. Following the selection procedure adopted in Ref.~\cite{Yang2014FermiBubbles}, models with halo heights $z=8~{\rm kpc}$ and $z=10~{\rm kpc}$ that fail to reproduce the $^{9}{\rm Be}/^{10}{\rm Be}$ data are discarded, and the remaining 64 GALPROP models are retained here as diffuse-emission templates. For each GALPROP run, we use the local interstellar spectra at $R=8.0~{\rm kpc}$ and $z=0$ as the spectrum at the Earth position. The all-electron background is then constructed by summing the primary electron, secondary electron, and secondary positron components in the GALPROP output. In Fig.~\ref{fig:eall}, the model spectra are compared with AMS-02\cite{Aguilar2014AMSElectron}, DAMPE\cite{Ambrosi2017DAMPE} and H.E.S.S.\cite{HESS2024CRE} results.

The comparison is performed by normalizing the GALPROP model spectra near $100~{\rm GeV}$ to the AMS-02 all-electron flux. This normalization removes the leading overall amplitude difference and is introduced only to facilitate a comparison of spectral shape at higher energies. We do not attempt to fit the all-electron spectrum over the full energy range shown in Fig.~\ref{fig:eall}. In particular, the spectrum below several tens of GeV is affected by solar modulation and by low-energy propagation effects such as reacceleration, convection, and the detailed low-energy injection spectrum. The low-energy hardening or curvature seen in precision lepton measurements is therefore a separate issue~\citep{Aguilar2019AMSElectrons,Strong2011ElectronSynchrotron,Potgieter2013SolarModulation}. In the present work we use the low-energy data only as context and focus our quantitative discussion on the high-energy regime above $\sim100~{\rm GeV}$, where radiative cooling limits the propagation horizon and where nearby-source or particle-physics-like TeV features are most relevant.

After this normalization, the GALPROP ensemble follows the broad trend of the AMS-02 and DAMPE all-electron data~\cite{Aguilar2014AMSElectron,Ambrosi2017DAMPE}. For clarity, Fig.~\ref{fig:eall} shows a representative subset of the retained GALPROP models rather than all 64 templates. These representative curves are selected to sample the spread of the normalized high-energy spectral shapes, in particular the variation between $100~{\rm GeV}$ and the TeV range. The colored curves therefore indicate the range of conventional diffuse-background shapes allowed by the selected propagation models.

The softening of the representative GALPROP backgrounds at multi-TeV energies is a consequence of the adopted conventional primary-electron source spectrum and propagation, not an additional local component. In the adopted GALPROP models, the primary-electron injection spectrum is parameterized as a three-segment broken power law in rigidity, with spectral indices $\gamma_{e,1}=1.6$, $\gamma_{e,2}\simeq2.4$, and $\gamma_{e,3}=4.0$, separated by break rigidities $R_{\rm br,1}\simeq2~{\rm GV}$ and $R_{\rm br,2}\simeq2.2~{\rm TV}$. This is a phenomenological but standard choice: the injected electron spectrum must be compatible with direct cosmic-ray lepton measurements, synchrotron constraints on the interstellar electron spectrum, and the fact that TeV electrons cool rapidly through synchrotron and inverse-Compton losses~\citep{Strong2007CRPropagation,Strong2011ElectronSynchrotron,Ackermann2012Diffuse}. The low-rigidity break describes the curvature required by low-energy electron and synchrotron constraints, whereas the high-rigidity break represents the decline of the primary-electron population at TeV energies. Synchrotron and inverse-Compton losses during propagation further reshape and broaden the high-rigidity break, producing a smooth softening in the propagated spectrum. Direct all-electron measurements also show that the spectrum is not a single power law up to the multi-TeV range, with H.E.S.S. and DAMPE reporting a high-energy softening or break~\citep{Aharonian2008HESS,Aharonian2009HESS,Ambrosi2017DAMPE}. 
An alternative interpretation was proposed by \citet{Shi2019CRE}, who attributed the electron--positron excess to continuously injecting sources distributed throughout the Galactic disk, with a pair-injection spectrum cutting off near $1~{\rm TeV}$ and a triple-power-law primary-electron spectrum. Although their source prescription differs from the GALPROP background adopted here, this work provides another example in which broad structure in the all-electron spectrum can arise from an astrophysical source population. Within the framework considered here,the break-like behavior in the GALPROP background should be understood as part of the conventional smooth diffuse-background model. It does not by itself represent a narrow nearby-source feature or a particle-physics-like endpoint.

For reference, we parameterize the median GALPROP spectrum with a smoothly broken power law,
$$
\Phi_{\rm bg}(E)=\Phi_0\left(\frac{E}{E_0}\right)^{-\gamma_1}\left[\frac{1+(E/E_b)^{1/s}}{1+(E_0/E_b)^{1/s}}\right]^{-s(\gamma_2-\gamma_1)},
$$
where we take $E_0=100~{\rm GeV}$ and fix $\Phi_0$ to the normalized median GALPROP flux at this energy. The indices $\gamma_1$ and $\gamma_2$ describe the asymptotic spectral slopes below and above the break, respectively, while $E_b$ gives the characteristic break energy and $s$ controls the smoothness of the transition. The fit gives $\gamma_1=3.12$, $\gamma_2=4.52$, $E_b=1.37~{\rm TeV}$, and $s=0.206$. It reproduces the median GALPROP spectrum with an rms fractional deviation of approximately $6\%$ over the fitted range. This parameterization is used only as a compact representation of the diffuse-background shape and does not introduce an additional physical component.

\begin{figure}
  \centering
  \includegraphics[width=0.95\linewidth]{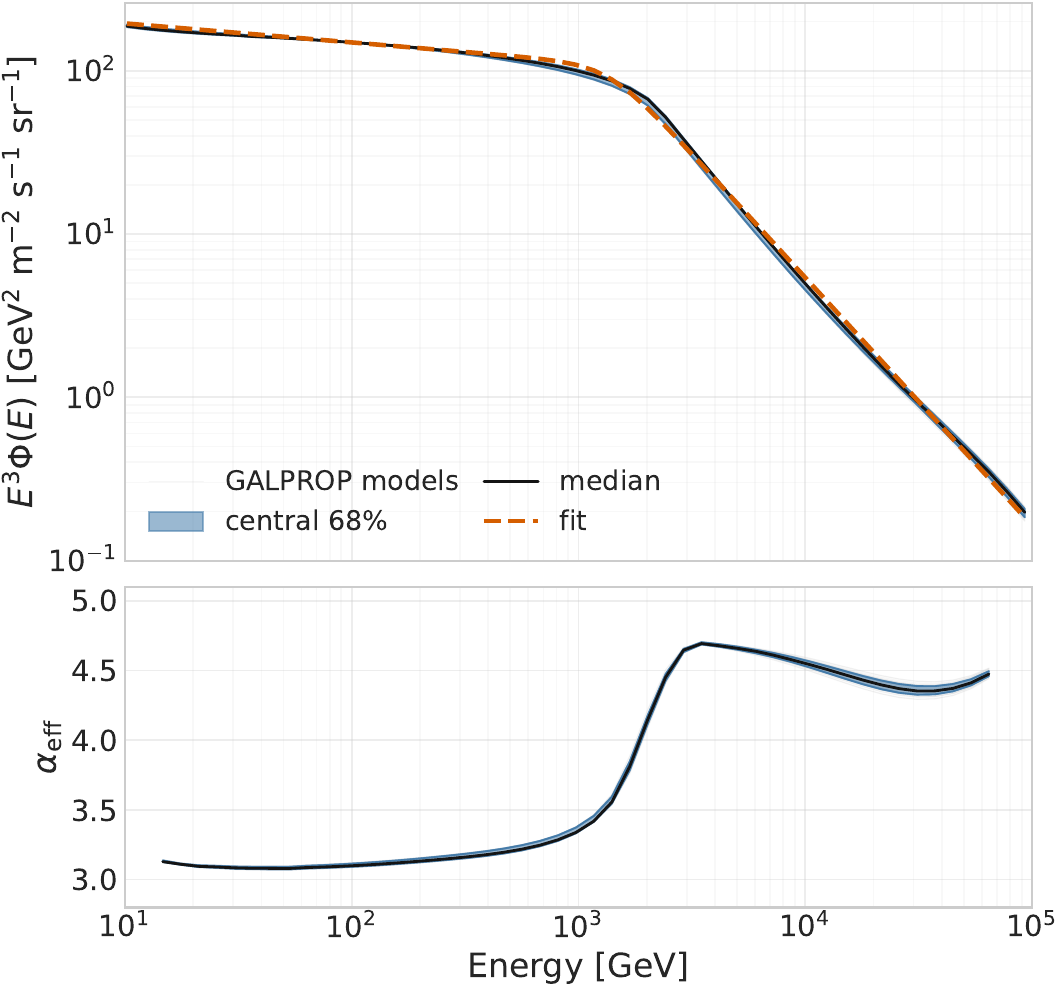}
  \caption{Diffuse all-electron backgrounds obtained from the GALPROP model ensemble. The upper panel shows the individual model spectra, the central $68\%$ interval, the median spectrum, and its smoothly broken power law fit. All spectra are normalized at $100~{\rm GeV}$. The lower panel presents the corresponding effective spectral indices, calculated directly from the individual GALPROP spectra.}
  \label{fig:galprop_alpha}
\end{figure}

Figure~\ref{fig:galprop_alpha} shows the GALPROP background ensemble and its energy-dependent spectral behavior. The upper panel presents the individual normalized model spectra, their central $68\%$ interval, the median spectrum, and the smoothly broken-power-law fit. The lower panel shows the effective spectral index,

$$
\alpha_{\rm eff}(E)=-\frac{d\ln\Phi_{\rm bg}(E)}{d\ln E},
$$
calculated directly from each GALPROP spectrum. The background is close to a power law below several hundred GeV, with a median index near $3.1$. It then softens smoothly across the TeV range and approaches an index of approximately $4.4$--$4.5$ at higher energies. Although the detailed curvature varies among the propagation models, the ensemble remains smooth and does not generate a narrow bump, an abrupt cutoff, or an edge-like structure.

%The smooth evolution of the GALPROP backgrounds provides the baseline for identifying additional high-energy components. A localized hardening, excess, or rapid change in the all-electron spectral slope relative to this background would not be naturally produced by the conventional diffuse component alone, and would indicate an additional contribution from either a nearby astrophysical source or a more exotic origin. We therefore consider nearby pulsars as a representative astrophysical scenario and use the Green-function solution for a continuously injecting point source to quantify the dependence of the observed spectrum on source age, distance, diffusion, and radiative cooling.

\section{Nearby pulsar contribution}
\label{sec:pulsar}

We next examine whether a nearby pulsar can produce a falling high-energy feature in the local all-electron spectrum. High-energy electrons and positrons from a localized source lose energy rapidly through synchrotron radiation and inverse-Compton scattering, so the observed spectrum depends sensitively on the source distance, age, and injection history. 
Since pulsars inject electron-positron pairs, the same source class can also contribute to the measured positron spectrum, as discussed in pulsar or pulsar wind nebula interpretations of the positron excess~\citep{Serpico2012PositronReview,Hooper2009Pulsars,Profumo2012Pulsars,Cholis2022PulsarPopulation}.
In this work, however, we do not perform a global charge-separated fit to the AMS-02 positron flux or positron fraction; the pulsar templates are used to study the shape and energetics of possible TeV-scale all-electron features.
We also do not attempt to model the antiproton spectrum. Antiprotons are expected to be dominated by secondary production in hadronic cosmic-ray interactions with interstellar gas during Galactic propagation~\cite{Strong2007CRPropagation,Moskalenko2001SecondaryAntiprotons,Aguilar2016AMSAntiproton}, whereas the pulsar templates considered here are leptonic $e^\pm$ sources. A simultaneous description of antiprotons would require a dedicated hadronic propagation calculation, including the primary proton and helium spectra, production cross sections, gas distribution, and solar modulation, and is outside the scope of the present all-electron spectral-shape study.
Following the Green-function treatment of electron propagation from discrete sources~\citep{Atoyan1995PRD,Aharonian1995AALetter,Kobayashi2004SNR}, we first write the solution for burst-like injection and then obtain the continuous-injection case by integrating over the source history. The injected spectrum is parameterized as
$$
Q(E_s)=Q_0E_s^{-\gamma}\exp\left[-\left(\frac{E_s}{E_{\rm cut}}\right)^\beta\right],
$$
where $E_s$ is the source energy, $\gamma$ is the injection index, $E_{\rm cut}$ is the cutoff energy, and $\beta$ controls the sharpness of the cutoff.

For the nearby-source calculation, we use a fixed diffusion-loss setup rather than scanning over the full GALPROP propagation ensemble. This allows us to isolate the dependence on source parameters. We take
$$
D(E)=D_0\left(\frac{E}{E_0}\right)^\delta,\qquad b(E)=b_0E^2,
$$
with $D_0=4.3\times10^{28}~{\rm cm^2~s^{-1}}$, $E_0=4~{\rm GeV}$, $\delta=0.415$, and $b_0=10^{-16}~{\rm GeV^{-1}~s^{-1}}$. Here $D(E)$ describes spatial diffusion and $b(E)$ approximates synchrotron and inverse-Compton cooling.

For an impulsive injection at time $t$ before observation, the propagated number density at distance $r$ is
$$
\psi_{\rm burst}(E,r,t)=Q(E_s)\frac{dE_s}{dE}G[\lambda^2(E,E_s),r],
$$
where
$$
E_s(E,t)=\frac{E}{1-b_0Et}
$$
is the injection energy required to observe an electron at energy $E$, and $dE_s/dE=(E_s/E)^2$ is the cooling Jacobian. The solution exists only for $b_0Et<1$. The diffusion kernel is
$$
G(\lambda^2,r)=\frac{1}{(\pi\lambda^2)^{3/2}}\exp\left(-\frac{r^2}{\lambda^2}\right),
$$
with
$$
\lambda^2(E,E_s)=4\int_E^{E_s}dE'\frac{D(E')}{b(E')}.
$$
The corresponding flux is $\Phi(E)=c\psi(E)/(4\pi)$.

The continuous-injection solution is obtained by summing the impulsive solution over the source lifetime,
$$
\psi_{\rm cont}(E,r,T)=\int_0^T dt\,Q[E_s(E,t)]\frac{dE_s}{dE}G[\lambda^2(E,E_s),r].
$$
Equivalently, changing variables from $t$ to $E_s$ gives
$$
\psi_{\rm cont}(E,r,T)=\frac{1}{b(E)}\int_E^{E_{\rm max}}dE_s\,Q(E_s)G[\lambda^2(E,E_s),r],
$$
where the upper limit is set by the finite source age and by the injected spectrum. Thus, burst-like and continuous injection are treated within the same Green-function framework: the former corresponds to a single propagation time, while the latter integrates over all injection times up to $T$. The characteristic cooling time is
$$
t_{\rm loss}(E)=\frac{1}{b_0E}\simeq3.2\times10^5~{\rm yr}\left(\frac{1~{\rm TeV}}{E}\right),
$$
so the finite-age cooling boundary and the diffusion kernel are the main propagation effects that reshape the source spectrum.

We consider two representative pulsar histories. The first is burst-like release from a mature nearby pulsar, an approximation commonly used for Geminga, Monogem, and other local sources relevant to the positron excess and the high-energy all-electron spectrum~\citep{Atoyan1995PRD,Hooper2009Pulsars,Yuksel2009Geminga,Malyshev2009PulsarsDM,Profumo2012Pulsars}. In this case most pairs are released over a time interval short compared with the propagation time. In the standard continuous-loss approximation, high-energy particles cool toward an age-dependent boundary,
$$
E_{\rm br}\simeq \frac{1}{b_0T}.
$$
If the intrinsic injection cutoff lies above the observed energy range, the turnover is mainly controlled by radiative cooling rather than by the intrinsic cutoff, and older sources peak at lower energies.

\begin{figure}
    \centering
    \includegraphics[width=1\linewidth]{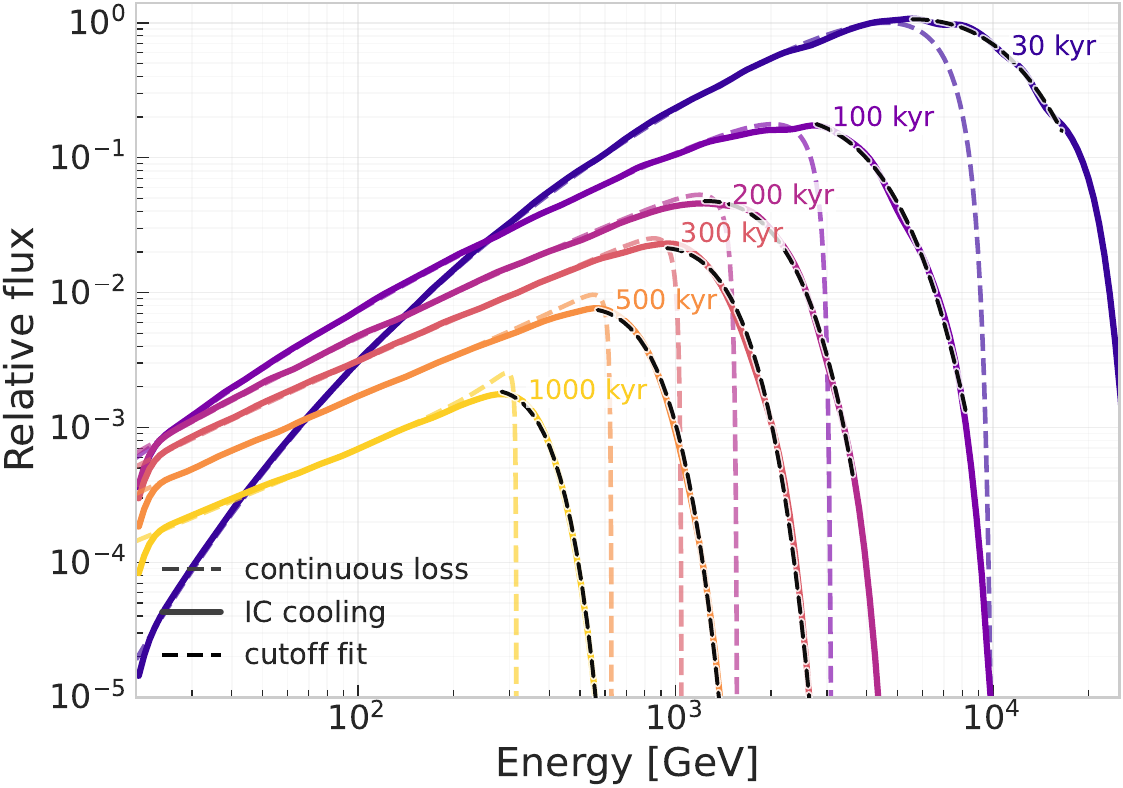}
    \caption{Burst-like propagated electron spectra for a source distance $d=500~{\rm pc}$. The injected spectrum is $Q(E_s)\propto E_s^{-1.5}\exp(-E_s/10~{\rm TeV})$, so the turnover is shaped jointly by radiative cooling and the source cutoff. Dashed curves show the standard continuous-loss calculation, while solid curves include stochastic inverse-Compton cooling using the path-averaged ISRF~\citep{Popescu2017ISRF} along the Geminga line of sight. All curves are computed with the same total injected electron energy and are rescaled by the maximum value among all curves in the figure. The black dashed curves show the empirical fits, $A E^{-p}\exp[-(E/E_{\rm eff})^{\beta_{\rm eff}}]$, performed over $E_{\rm pk}\leq E\leq 3E_{\rm pk}$ with $p$ fixed to the corresponding injection index.}
    \label{fig:pulsar_burst_cooling}
\end{figure}

Figure~\ref{fig:pulsar_burst_cooling} shows this burst-like case for $d=500~{\rm pc}$ and source ages from $30~{\rm kyr}$ to $1000~{\rm kyr}$. The dashed curves show the standard continuous-loss result, in which the spectrum develops a pronounced peak followed by a sharp decline near the cooling boundary. This sharp cutoff should be interpreted with care. For TeV electrons, inverse-Compton scattering is affected by Klein--Nishina effects, and individual scatterings can remove a non-negligible fraction of the electron energy. The cooling history is therefore stochastic rather than perfectly deterministic, as emphasized by \citet{JohnLinden2023NoSharpFeatures}. The finite injection cutoff at $10~{\rm TeV}$ suppresses the formal pile-up expected for an unbounded hard injection spectrum with $\gamma<2$; the turnover in Fig.~\ref{fig:pulsar_burst_cooling} is therefore shaped jointly by radiative cooling and the source cutoff, with cooling becoming more important for older sources.

To estimate this effect, we implement a stochastic inverse-Compton Monte Carlo using the Galactic interstellar radiation field (ISRF) model of \citet{Popescu2017ISRF}. We take the Geminga line of sight, $(l,b)=(195.1^\circ,4.3^\circ)$, and average the tabulated ISRF spectrum from the Solar position to $d=500~{\rm pc}$. The resulting path-averaged energy densities are $\rho_{\rm opt}=0.435~{\rm eV~cm^{-3}}$, $\rho_{\rm IR}=0.504~{\rm eV~cm^{-3}}$, and $\rho_{\rm CMB}=0.260~{\rm eV~cm^{-3}}$. The optical and infrared components are taken directly from the tabulated spectral energy density, rather than approximated as single-temperature blackbodies, and are converted into the photon number density used in the Klein--Nishina kernel. For each injected electron, we sample individual inverse-Compton scatterings with the isotropic Klein--Nishina kernel, while we assume a uniform magnetic field of \(B=3~\mu{\rm G}\), corresponding to \(U_B\simeq0.225~{\rm eV~cm^{-3}}\), and treat synchrotron cooling as a continuous loss process. Each Monte Carlo trajectory is assigned a diffusion weight using $\lambda_{\rm MC}^2=4\int_0^T D[E(t)]\,dt$, which reduces to the standard diffusion length in the continuous-loss limit. 

For each source age, the stochastic spectra are generated from $10^5$ Monte Carlo trajectories drawn from the injected spectrum, rather than from a fixed number of particles in each final-energy bin. The same set of trajectories is used both to construct the stochastic spectra shown in Fig.~\ref{fig:pulsar_burst_cooling} and to obtain the empirical $\beta_{\rm eff}$ values in Table~\ref{tab:effective_cutoff_beta}. The final energies are accumulated in 189 logarithmic bins between $20~{\rm GeV}$ and $30~{\rm TeV}$. 
To reduce bin-to-bin fluctuations, the weighted histograms are convolved with a Gaussian kernel with a fixed standard deviation of $2.4$ energy bins, corresponding to $\sigma_{\log_{10}E}\simeq0.040$. The same smoothing prescription is applied to all source ages. The quoted bootstrap uncertainties describe Monte Carlo statistical fluctuations for this fixed analysis procedure and do not include systematic variations associated with the binning, smoothing width, or fitting window. Although these choices affect the precise fitted values of $\beta_{\rm eff}$, they do not alter the conclusion that stochastic inverse-Compton cooling broadens the sharp edge obtained with deterministic continuous losses.
We also tested the Monte Carlo convergence using repeated random subsamples containing $5\times10^3$--$5\times10^4$ trajectories. The results obtained with $5\times10^4$ trajectories are statistically consistent with the $10^5$-trajectory calculation for all source ages, and we therefore adopt $10^5$ trajectories for each age.

The solid curves in Fig.~\ref{fig:pulsar_burst_cooling} preserve the broad age dependence but exhibit a smoother post-peak decline and a more extended high-energy tail than the conventional $b_0E^2$ continuous-loss baseline. The dashed and solid curves use the same source and diffusion parameters, but their energy-loss prescriptions are not identical: the dashed curves adopt the effective deterministic loss law $b(E)=b_0E^2$, whereas the Monte Carlo calculation treats synchrotron cooling continuously and samples individual inverse-Compton scatterings from the \citet{Popescu2017ISRF} ISRF using the Klein--Nishina kernel. The quantitative difference between the two calculations therefore reflects both the stochastic nature of inverse-Compton scattering and the energy dependence of the Klein--Nishina loss rate. Nevertheless, the broader high-energy decline is consistent with the behavior identified by \citet{JohnLinden2023NoSharpFeatures} and shows that the narrow edge obtained with the conventional continuous-loss approximation is not robust under a more realistic cooling treatment.

The second case is a young nearby pulsar that continuously injects pairs over its finite lifetime. Here the falling feature is mainly inherited from the source spectrum itself. We adopt an intentionally hard and sharply cut off injection spectrum,
$
Q(E_s)\propto E_s^{-1}\exp[-(E_s/500~{\rm GeV})^2],
$
at a fixed distance $d=100~{\rm pc}$. Hard pair spectra with high-energy cutoffs are commonly used in pulsar interpretations of electron and positron data~\citep{Aharonian1995AALetter,ZhangCheng2001,Hooper2009Pulsars,Profumo2012Pulsars}. Our choice is deliberately extreme, designed to test how sharp a continuous-injection pulsar spectrum can become under favorable source assumptions.

\begin{figure}
    \centering
    \includegraphics[width=1\linewidth]{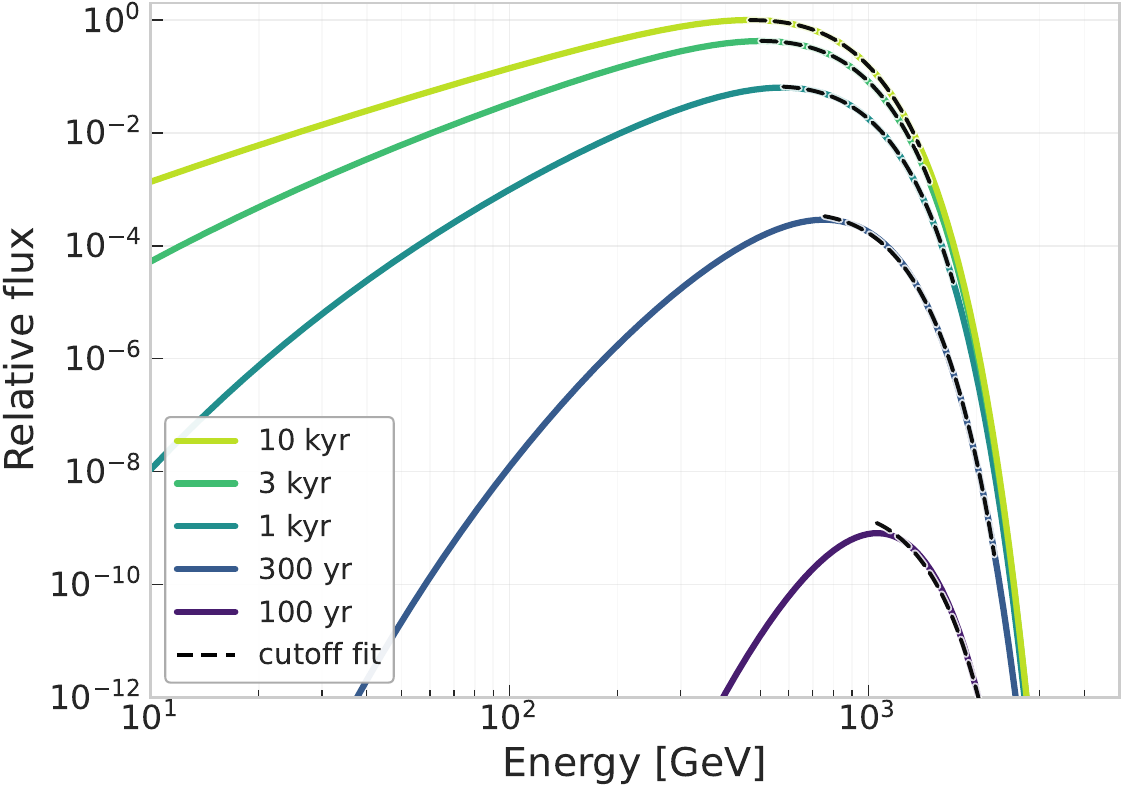}
    \caption{Propagated electron spectra for the continuous-injection intrinsic-cutoff case at a fixed source distance of $d=100~{\rm pc}$. The injected spectrum is $Q(E_s)\propto E_s^{-1}\exp[-(E_s/500~{\rm GeV})^2]$, and different curves correspond to source ages from $100~{\rm yr}$ to $10~{\rm kyr}$. All curves are computed with the same injection normalization and are rescaled by the maximum value among all curves in the figure, so the relative flux differences between different ages are retained. The black dashed curves show the empirical fits, $A E^{-p}\exp[-(E/E_{\rm eff})^{\beta_{\rm eff}}]$, performed over $E_{\rm pk}\leq E\leq 3E_{\rm pk}$ with $p$ fixed to the corresponding injection index.}
    \label{fig:pulsar_intrinsic_cutoff}
\end{figure}

Figure~\ref{fig:pulsar_intrinsic_cutoff} shows the propagated spectra for source ages from $100~{\rm yr}$ to $10~{\rm kyr}$. For the youngest sources, the arriving flux is strongly suppressed because the injection time is short and particles have limited time to diffuse to Earth. As the source age increases, the accumulated injection grows and the peak shifts to lower energy. Unlike the burst-like cooling feature, the rapid falloff above the peak is mostly the propagated tail of the intrinsic super-exponential cutoff, mildly shifted and broadened by diffusion and radiative cooling.

To compare the apparent sharpness of these spectra, we fit the peak and post-peak falling side with the empirical form
$A E^{-p}\exp[-(E/E_{\rm eff})^{\beta_{\rm eff}}]$.
We fix $p$ to the injection index of the corresponding source model: $p=1.5$ for the burst-like pulsar spectra and $p=1.0$ for the continuous-injection intrinsic-cutoff spectra. The fit starts from the peak energy of $E^3\Phi(E)$. For the stochastic inverse-Compton and continuous-injection intrinsic-cutoff spectra, the fit extends from $E_{\rm pk}$ to $3E_{\rm pk}$. Fixing $p$ avoids the strong degeneracy among $p$, $E_{\rm eff}$, and $\beta_{\rm eff}$ in a narrow post-peak fitting window. The fitted $\beta_{\rm eff}$ values are therefore used only as diagnostic measures of the apparent propagated spectral sharpness, not as precision physical parameters. For the stochastic inverse-Compton spectra, the same $10^5$ Monte Carlo trajectories used to construct Fig.~\ref{fig:pulsar_burst_cooling} are used to obtain the empirical $\beta_{\rm eff}$ values in Table~\ref{tab:effective_cutoff_beta}, and the quoted uncertainties are Monte Carlo statistical errors from bootstrap resampling only. Once stochastic inverse-Compton cooling is included, the sharp burst-like cooling boundary is broadened and the effective post-peak sharpness is reduced to values of order a few. By contrast, the continuous-injection case gives $\beta_{\rm eff}\simeq2$, showing that its sharpness is mainly inherited from the assumed super-exponential intrinsic cutoff rather than generated by propagation.

\begin{table}
\centering
\caption{Empirical cutoff-fit parameters for the propagated local-source spectra. The spectra are fitted with $A E^{-p}\exp[-(E/E_{\rm eff})^{\beta_{\rm eff}}]$, with $p$ specified in each group heading. The peak energy $E_{\rm pk}$ is defined as the energy at which $E^3\Phi(E)$ reaches its maximum. For stochastic inverse-Compton rows, the uncertainties on $\beta_{\rm eff}$ are bootstrap $1\sigma$ Monte Carlo statistical errors from 200 resamplings of the same $10^5$ trajectories used for Fig.~\ref{fig:pulsar_burst_cooling}. The parameters describe only the apparent propagated spectral shape and should not be identified with the intrinsic injection parameters.}
\label{tab:effective_cutoff_beta}
\setlength{\tabcolsep}{7pt}
\renewcommand{\arraystretch}{1.12}
\begin{tabular}{cccc}
\hline
$T$ [kyr] & $E_{\rm pk}$ [TeV] & $E_{\rm eff}$ [TeV] & $\beta_{\rm eff}$ \\
\hline
\multicolumn{4}{c}{Burst-like pulsar, stochastic IC ($p=1.5$)} \\
\hline
30   & 5.57  & 4.97  & $1.29\pm0.31$ \\
100  & 2.78  & 2.15  & $1.56\pm0.39$ \\
200  & 1.23  & 1.48  & $2.17\pm0.21$ \\
300  & 0.94 & 1.11  & $2.65\pm0.21$ \\
500  & 0.57 & 0.65 & $2.89\pm0.15$ \\
1000 & 0.28 & 0.33 & $3.61\pm0.14$ \\
\hline
\multicolumn{4}{c}{Continuous-injection pulsar ($p=1.0$)} \\
\hline
0.1 & 1.06  & 0.80 & 2.54 \\
0.3 & 0.75 & 0.67 & 2.36 \\
1   & 0.58 & 0.56 & 2.15 \\
3   & 0.50 & 0.50 & 2.02 \\
10  & 0.47 & 0.47 & 1.95 \\
\hline
\multicolumn{4}{c}{SNR-like burst source ($p=2.4$)} \\
\hline
10 & 7.11 & 13.9 & 2.72 \\
\hline

\end{tabular}
\end{table}

We next compare these spectral shapes with the required source energetics. For each case, we rescale the pulsar contribution to reach $10\%$ of the median diffuse all-electron background at a reference energy. For burst-like mature pulsars, the relevant quantity is the total injected pair energy. Taking $d=500~{\rm pc}$ and normalizing at $1~{\rm TeV}$, we find $W_e\simeq5\times10^{47}~{\rm erg}$ for $T=100~{\rm kyr}$ and $W_e\simeq5\times10^{48}~{\rm erg}$ for $T=300~{\rm kyr}$. The increase with age is expected because older electrons have cooled more strongly at TeV energies. These values are within the broad range used in pulsar interpretations of the positron excess, where escaping pair energies of $10^{47}$--$10^{48}~{\rm erg}$ are typical and values approaching $10^{49}~{\rm erg}$ require very efficient or energetic sources~\citep{Hooper2009Pulsars,HallHooper2009ATIC,Malyshev2009PulsarsDM}. Thus, the $100~{\rm kyr}$ case is energetically reasonable under favorable assumptions, while the $300~{\rm kyr}$ case is already demanding.

For young continuous injection, we normalize at $500~{\rm GeV}$, close to the intrinsic cutoff energy. For $T=10~{\rm kyr}$ and $d=100~{\rm pc}$, reaching $10\%$ of the diffuse background requires $\dot W_e\simeq1.1\times10^{34}~{\rm erg~s^{-1}}$, or $W_e\simeq3.4\times10^{45}~{\rm erg}$ over the source lifetime. This is possible for an efficient nearby pulsar. For $T=300~{\rm yr}$ at the same distance, however, the required power rises to $\dot W_e\simeq7.3\times10^{37}~{\rm erg~s^{-1}}$, corresponding to $W_e\simeq6.9\times10^{47}~{\rm erg}$. This is comparable to the spin-down power of the most powerful young pulsars and is difficult to realize once only a fraction of the spin-down power can escape as high-energy pairs. Energetic pulsars typically have $\dot E\sim10^{35}$--$10^{37}~{\rm erg~s^{-1}}$, with only the most powerful systems reaching $\sim10^{38}~{\rm erg~s^{-1}}$~\citep{Manchester2005ATNF,Gaensler2006PWNReview,Kargaltsev2008PWNe}; population studies usually require pair-conversion efficiencies of several percent to tens of percent~\citep{Cholis2022PulsarPopulation,Bitter2022PulsarPopulation}. Therefore, the continuous-injection scenario is plausible for a nearby $\sim10~{\rm kyr}$ source, but not for a very young source with a short injection history.

These results show that nearby pulsars can produce structured high-energy electron spectra under favorable conditions. However, a rapid endpoint-like falloff requires additional assumptions. In the burst-like case, the apparent cooling edge is softened once stochastic inverse-Compton losses are included. In the young continuous-injection case, the rapid decline is mainly inherited from an intrinsic cutoff already present in the injected pair spectrum. Therefore, an observable pulsar-induced feature must be assessed together with the smooth diffuse background, the source age and distance, and the required pair energy budget.

\section{A possible new component above TeV energies}
\label{sec:tev_component}

\begin{figure}
    \centering
    \includegraphics[width=1\linewidth]{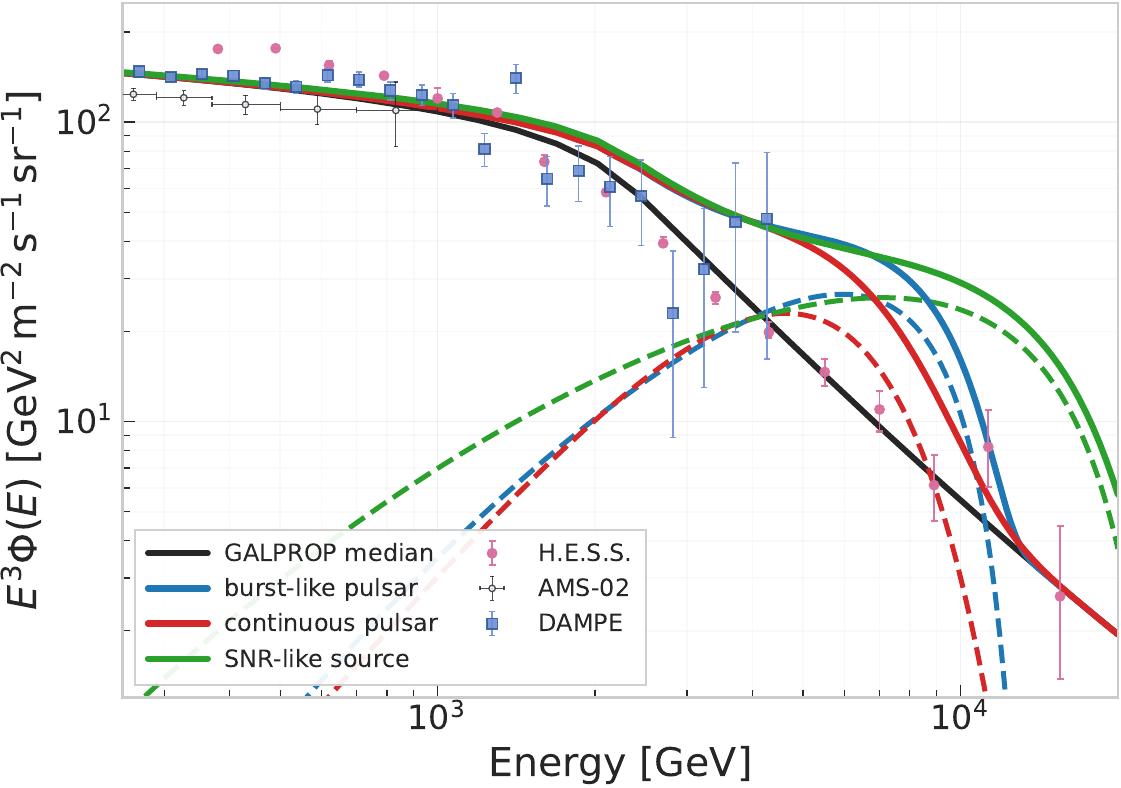}
    \caption{Illustrative local-source contributions normalized to the positive residuals of the two highest-energy DAMPE points relative to the median GALPROP diffuse background. Dashed curves show the source components alone, while solid colored curves show the sum of the GALPROP background and the corresponding source contribution. The three examples correspond to a burst-like pulsar, a continuous-injection pulsar, and an SNR-like burst source. Here we do not use the H.E.S.S. points to normalize the source templates.}
    \label{fig:dampe_last_two_pulsar_fit}
\end{figure}

The DAMPE all-electron spectrum shows a clear softening near the TeV scale and has motivated extensive discussions of possible additional high-energy components~\citep{Ambrosi2017DAMPE,Fowlie2018DAMPESquib,Yuan2017DAMPEInterpretations,Huang2018SharpStructures}. In the present data, the two highest-energy DAMPE points have large statistical uncertainties and are still compatible with the diffuse-background expectation within their error bars. Nevertheless, their central values lie mildly above the median GALPROP background. If this positive residual reflects a real component rather than a statistical fluctuation, its origin would naturally be local: TeV electrons cool rapidly through synchrotron radiation and inverse-Compton scattering, so only sources within a limited distance and time interval can contribute efficiently. Nearby pulsars are one possibility, as discussed above, while nearby supernova remnants (SNRs) provide another well-known class of discrete electron sources. In particular, \citet{Kobayashi2004SNR} showed that nearby remnants can imprint identifiable, source-dependent structures in the $1$--$10~{\rm TeV}$ electron spectrum. We therefore use the two highest-energy DAMPE points as an illustrative estimate of what kind of local source component would be required, without interpreting them as statistically significant evidence for a new component. 
The recent H.E.S.S. spectrum \cite{HESS2024CRE} is also shown in Fig.~\ref{fig:dampe_last_two_pulsar_fit} for comparison. Since it is inferred indirectly from air-shower measurements and is subject to appreciable systematic uncertainties and residual hadronic contamination, especially above a few TeV, we do not use the H.E.S.S. points to normalize the source templates.

Fig.~\ref{fig:dampe_last_two_pulsar_fit} shows three illustrative local-source templates. In each case, the source component is normalized to the residual relative to the median GALPROP background, so that the total background-plus-source curve matches the central values of the two highest-energy DAMPE points. 
The burst-like pulsar example uses the same injection spectrum as in Sec.~\ref{sec:pulsar}, with $d=500~{\rm pc}$ and an illustrative age of $T\simeq20~{\rm kyr}$, requiring $W_e\simeq8\times10^{46}~{\rm erg}$. The continuous-injection pulsar example adopts a favorable TeV-scale intrinsic cutoff, with $d=100~{\rm pc}$, $T=3~{\rm kyr}$, and $E_{\rm cut}=5~{\rm TeV}$, giving $\dot W_e\simeq6\times10^{33}~{\rm erg~s^{-1}}$ and $W_e\simeq5\times10^{44}~{\rm erg}$. For comparison, the SNR-like burst template follows the Kobayashi-type setup with $\gamma=2.4$, $E_{\rm cut}=20~{\rm TeV}$, $d=300~{\rm pc}$, and gives a representative requirement of $W_e\simeq4\times10^{47}~{\rm erg}$ for a young source with $T\simeq10~{\rm kyr}$.
Because this SNR-like template adopts a softer injected electron spectrum than the pulsar examples, and only a simple exponential cutoff rather than a sharp or super-exponential intrinsic cutoff, its propagated contribution is broader and less endpoint-like. A fit of the local post-peak region with $A E^{-p}\exp[-(E/E_{\rm eff})^{\beta_{\rm eff}}]$, fixing $p=2.4$, gives $\beta_{\rm eff}\simeq2.7$ for the Fig.~\ref{fig:dampe_last_two_pulsar_fit} SNR-like component.

These estimates are not intended as evidence for a statistically significant excess. They simply show that a mild multi-TeV residual, if confirmed, could be described by a nearby discrete source with energetics that depend strongly on the source class, distance, age, and injection history. Pulsar scenarios can produce relatively narrow high-energy structures under favorable cutoff or cooling conditions, while the SNR-like template considered here gives broader TeV-scale contributions. Improved multi-TeV electron measurements will therefore be needed to determine whether the high-energy spectrum is consistent with a smooth diffuse background or contains an additional local component.

%69
\section{Discussion and outlook}
\label{sec:discussion}

The interpretation of TeV-scale structure in the cosmic-ray all-electron spectrum depends critically on the separation between the smooth diffuse background and any additional local component. In this work we used GALPROP models calibrated by cosmic-ray nuclei and diffuse gamma-ray observations to characterize the conventional all-electron background. The resulting spectra remain smooth above $\sim100~{\rm GeV}$ and gradually soften toward higher energies. 

%They do not generate a localized TeV-scale hardening, cutoff, or edge-like structure by themselves. Therefore, if a sufficiently sharp feature is confirmed in the all-electron spectrum, it should be treated as evidence for an additional component rather than as a natural consequence of the conventional diffuse background alone.

We then examined whether nearby pulsars can provide such an additional component. A young continuously injecting pulsar with an intrinsic super-exponential cutoff can produce a rapidly falling spectrum, but the sharpness is inherited mainly from the assumed source cutoff and the propagated cutoff is mildly broadened rather than sharpened. The energetics are plausible for a very nearby source with age of order $10~{\rm kyr}$, but become severe for a few-hundred-year source because the short injection history strongly suppresses the arriving flux. A burst-like mature pulsar provides a complementary mechanism: if the injection cutoff is well above the TeV range, the observed turnover can be controlled by the cooling break and can shift with source age. For Geminga-like ages, the required total injected pair energy can lie in the range commonly invoked for pulsar interpretations of the positron excess, but the sharpness of the cooling feature depends on idealized assumptions about impulsive release and continuous radiative losses. The numerical investigation shows that, in these cases, the effective post-break sharpness parameter $\beta_{\rm eff}$ remains moderate and varies with source age and the adopted fitting prescription. Across the source ages and loss prescriptions considered here, the spectra evolve continuously rather than producing a stable, source-age-independent sharp edge.

If the observed structure is a smooth cutoff or a cooling-broadened decline, it would remain naturally compatible with nearby astrophysical sources. In contrast, a very sharp and stable edge-like feature would be more difficult to accommodate with ordinary pulsar propagation and would strengthen the case for alternative origins.

Future high-energy-resolution observations will provide a more precise measurement of the cosmic-ray all-electron spectrum.
With increasing DAMPE exposure and future observations by HERD\cite{Zhang2014HERD}, the TeV-scale all-electron spectrum can be measured with improved statistics and energy resolution. Such data can determine whether a putative feature is a smooth astrophysical cutoff, a cooling-broadened local-source contribution, or a sharper edge-like structure more naturally associated with particle physics. In this sense, the all-electron spectrum provides not only a probe of local cosmic-ray sources, but also a diagnostic of whether an observed high-energy structure has a conventional astrophysical origin or points to new particle physics.

\section{Acknowledgment}
Ruizhi Yang is supported by the NSFC under grants
12588101, 12393854, and by the Natural Science Foundation of
Sichuan Province under grant 2025ZNSFSC0065. Ruizhi Yang
gratefully acknowledges the support of Cyrus Chun Ying Tang
Foundations and of the studio of Academician Zhao Zhengguo,
Deep Space Exploration Laboratory.

\bibliography{main}% Produces the bibliography via BibTeX.

\end{document}